\documentclass[apjl]{emulateapj}
\usepackage{graphicx}

\def\thalf{{\textstyle{1 \over 2}}}

\def\lexp{\mathop{\langle}\nolimits}
\def\rexp{\mathop{\rangle}\nolimits}

\def\del{\nabla}
\def\rhobar{{\bar \rho}}
\def\adot{{\dot a}}
\def\lin{{\rm lin}}
\def\min{{\rm min}}
\def\x{{x}}

\begin{document}

\title{Effects of Inhomogeneities on Cosmic Expansion}

\author{E. R. Siegel and J. N. Fry}

\affil{Department of Physics, University of Florida, Gainesville, 
FL 32611-8440; \break siegel@phys.ufl.edu, fry@phys.ufl.edu}

\begin{abstract}

We evaluate the effect of inhomogeneity energy on the expansion 
rate of the universe.  Our method is to expand 
to Newtonian order in potential and velocity 
but to take into account fully nonlinear density inhomogeneities.
To linear order in density, kinetic and gravitational potential 
energy contribute to the total energy of the universe with the 
same scaling with expansion factor as spatial curvature.
In the strongly nonlinear regime, growth saturates, and the net 
effect of the inhomogeneity energy on the expansion rate remains 
negligible at all times.
In particular, inhomogeneity contributions never mimic the effects 
of dark energy or induce an accelerated expansion.

\end{abstract}

\subjectheadings{cosmological parameters --- 
cosmology : theory --- large-scale structure of universe}

\section{Introduction}

Recent observations of type-Ia supernovae \citep{Ries00} and 
the cosmic microwave background \citep{Benn03} in tandem 
suggest that the cosmological expansion is accelerating.
Understanding the source of this accelerated expansion is one of 
the greatest current unsolved problems in cosmology \citep{Peeb03}.
Acceleration seems to render inadequate a universe consisting entirely 
of matter, and appears to require an additional, unknown type of energy 
(dark energy, perhaps realized as a cosmological constant).
An alternative to dark energy is that acceleration arises from 
a known component of the universe whose effects 
on the cosmic expansion have not been fully examined.
One possibility currently being examined is that inhomogeneities 
in a matter dominated universe, on either sub-horizon \citep{Ras05,Nota05} 
or super-horizon scales \citep{Kolb04,Kolb05a,Kolb05b,Bara05}, 
may influence the expansion rate at late times.
The central idea is that the energy induced by inhomogeneities 
leads to additional source terms in the Friedmann equations, 
with effects on the dynamics that leave no need for a separate 
dark energy component.
In their entirety, these proposals present conflicting claims and 
a general state of much confusion: 
does the inhomogeneity energy produce an accelerated expansion, 
acting in effect as dark energy \citep{Kolb05b}, 
or does it behave as curvature \citep{Gesh05}?
Is the magnitude of the effect small, large, or even divergent, 
on either large scales \citep{Kolb05b}, 
or on small scales at late times \citep{Nota05}?

Part of the confusion arises from the fully relativistic 
perturbation theory formulation of many of these calculations.
Although this is undeniably a valid approach, the number of terms 
in a perturbation theory calculation can be large 
and can mask the underlying physics.
In this {\it Letter}, taking advantage of phenomenological results 
that have been derived from a combination of quasilinear 
perturbation theory, nonlinear theory, and numerical simulations, 
we compute the potential and kinetic inhomogeneity energies 
within the horizon to Newtonian order in potential and velocity 
for fully nonlinear density contrasts.
We find these energies to be small at present, and their projected values 
remain small, even far into the future.
Section 2 considers the effect of inhomogeneities, 
for weak gravity and slow motions but for arbitrary 
density perturbations, 
characterized in terms of the density power spectrum.
Section 3 presents the results for the kinetic and potential 
energies in both the linear and the fully nonlinear regimes, 
as a function of the cosmological expansion factor.
Finally, section 4 discusses the implications of these results. 
%for the present and future expansion history of the universe.

\section{Effects of Inhomogeneities}

The purpose of our work is to investigate whether inhomogeneity energy 
can mimic the effects of dark energy for a universe containing only matter.
To this end, we work in an $\Omega_m = 1$ Einstein-de Sitter universe, 
with no curvature or cosmological constant, and compute the effects of 
inhomogeneities on the cosmic expansion rate.
The dynamics of cosmological expansion are governed by the Friedmann 
equations, 
\begin{equation} 
\Bigl( {\dot a \over a} \Bigr)^2 = {8 \pi \over 3} \, G \rho , \qquad 
\Bigl( {\ddot a \over a} \Bigr) = -{4 \pi \over 3} \, G (\rho + 3 p) .
\label{Friedman}
\end{equation}
Any mass or energy density that makes up a significant fraction  
of the total can influence the evolution of the cosmological 
scale factor $ a(t) $.
A contribution to the energy density of the universe with equation 
of state $ p_i = w \rho_i $ has $ \rho_i \propto a^{-3(1+w)} $, 
or $ \rho_i / \rho_m \propto a^{-3w} $; 
in particular, a component with $ \rho \propto a^{-2} $ behaves 
as $ w = -{1 \over 3} $ or curvature, and a component 
with constant $ \rho $ behaves as a cosmological constant or dark energy.

We introduce the effects of inhomogeneities following the formulation 
of \citet{SH96}.
In the conformal Newtonian gauge, with metric 
\begin{equation} 
ds^2 = a^2(\tau) [-(1 + 2 \psi) d\tau^2 + (1 - 2 \phi) d\x^2] ,
\end{equation} 
the time-time Einstein equation ($G^0_{\, \, 0}$) yields 
\begin{eqnarray}
3 \Bigl( {\dot a \over a} \Bigr)^2 (1 - 2 \psi) + (2 &+& 6 \phi) 
{1 \over a^2} \, \del^2\!\phi + {1 \over a^2} (\del \phi)^2 
\nonumber \\ 
&=& 8 \pi G \rhobar (1 + \delta) (1 + v^2) \mathrm{,}
\end{eqnarray} 
Where $\phi \simeq \psi$ from the space-space components of 
$G^\mu_{\, \, \nu}$.  (Our numerical factors are slightly different from 
those of Seljak \& Hui; they make little difference in the results.)
The source on the right-hand side includes a density perturbation 
$ \delta = \delta \rho / \rhobar $ in the material rest frame, 
with the transformation to the cosmological frame expanded to leading 
order for small $v^2$.
Ignoring $ \phi \, \del^2 \!\phi $, $ (\del \phi)^2 $, and $ v^2 $, the 
homogeneous part of this equation reproduces the usual Friedmann equation.  
The inhomogeneous part reveals that $\phi$ obeys the Poisson equation 
with source $ 4 \pi G \rhobar a^2 \delta $.
The volume average of the entire equation then leads to 
\begin{equation} 
\Bigl( {\dot a \over a} \Bigr)^2 = {8 \pi \over 3} G \rhobar \, 
\left( 1 - 5 \, W + 2K \right) , 
\end{equation}
where $W$ and $K$ are the Newtonian potential and kinetic 
energy per unit mass,  
\begin{equation} \label{wk}
W = \thalf \lexp (1+\delta) \, \phi \rexp , \qquad 
K = \thalf \lexp (1 + \delta) v^2 \rexp .
\end{equation}
These expressions are correct to first order in $\phi$ and $v^2$, 
but neither an assumption nor an approximation in $\delta$.
We assume that $ \lexp \del^2 \phi \rexp = 0 $; in all other places 
the Poisson equation is adequate to determine $\phi$.

The Newtonian potential and kinetic energies thus can 
influence cosmological expansion.
We can compute both $W$ and also $K$ completely 
and exactly from knowledge only  of the density power spectrum.
The potential is related to the density inhomogeneity by the 
Poisson equation, $ \del^2 \phi = 4 \pi G \rhobar a^2 \, \delta $, 
an expression which holds even for nonlinear inhomogeneities.
From this, we obtain 
\begin{equation} \label{W}
W = -{1 \over 2} \, 4 \pi G \rhobar a^2 \int {d^3k \over (2\pi)^3} 
\, {P(k) \over k^2} = -\int {dk \over k} \, \Delta_W^2(k) , 
\end{equation}
an expression correct in both linear and nonlinear regimes if $ P(k) $ 
is the appropriate linear or nonlinear power spectrum.
The last equality defines the dimensionless spectral density 
$ \Delta_W^2(k) $.

In linear perturbation theory, valid for small inhomogeneities, 
the density contrast grows as $ \delta = \delta_0(\x) D(t) $, 
where in a matter dominated universe 
$ D(t) \propto a(t) \propto t^{2/3} $ \citep{Peeb80}.
The kinetic energy follows from the linearized equation of continuity, 
$ \dot \delta + \del \cdot v/a = 0 $ \citep{Peeb80}, 
\begin{equation} \label{Klin}
K_\lin = {1 \over 2} \,  \adot^2 
\int {d^3k \over (2\pi)^3} \, {P(k) \over k^2} 
\end{equation}
(the usual factor $ f(\Omega) \simeq \Omega^{0.6} = 1 $ for $ \Omega_m = 1 $).
The kinetic energy scales with $ a(t)$ as $ \adot^2 D^2 $, 
while the potential energy scales as $ \rhobar a^2 D^2 $; 
and so both $W$ and $K$ grow as $ D^2/a \propto a(t) $, 
or $ \rho_U = \rhobar (W+K) \propto a^{-2} $.
As was noted by \citet{Gesh05} for super-horizon inhomogeneities, 
in perturbation theory inhomogeneity energy has the same effect 
on the expansion rate as spatial curvature.
We note that $ K_\lin/|W_\lin| = H^2 / 4\pi G \rhobar = {2 \over 3} $, 
a fixed ratio in the linear regime.
The full kinetic energy in principle involves higher order correlation 
functions and is not a simple integral over the power spectrum.
Nonetheless, the full kinetic energy can be obtained simply 
from the potential energy through the cosmic energy equation 
of \citet{Irv61} and \citet{Layz63}, 
\begin{equation} \label{CEE}
\left(\frac{d}{dt} + \frac{2\adot}{a}\right)K =
- \left(\frac{d}{dt} + \frac{\adot}{a}\right)W , 
\end{equation}
with initial conditions set in the linear regime, 
$ K_\lin = {2 \over 3} |W_\lin| $.
Equations (\ref{W}) and (\ref{CEE}) provide us with expressions
sufficient to calculate nonperturbative contributions to the 
expansion rate for both the gravitational potential perturbation 
and kinetic energy components.  The results of these calculations
are given in the next section.

\section{Results}

Equations~(\ref{W}) and (\ref{CEE}) determine the 
inhomogeneity energy of the universe as a function of epoch, 
which we characterize by the expansion factor $ a/a_0 $.
For the primordial power spectrum, we use the CDM power 
spectrum as given by \citet{BBKS86}, 
with spectral index $ n = 1 $, $ \Omega_m = 1 $, 
and COBE normalized amplitude 
$ \delta_H = 1.9 \times 10^{-5} $ \citep{BW97}.
%The BBKS form of the power spectrum does not contain such details 
%as baryon oscillations, but is accurate enough for this purpose.
To obtain the nonlinear power spectrum we use the linear-nonlinear 
mapping of \citet{PD94,PD96}.
The results of these calculations are shown in Figures 1 and 2.

\includegraphics[width=\columnwidth]{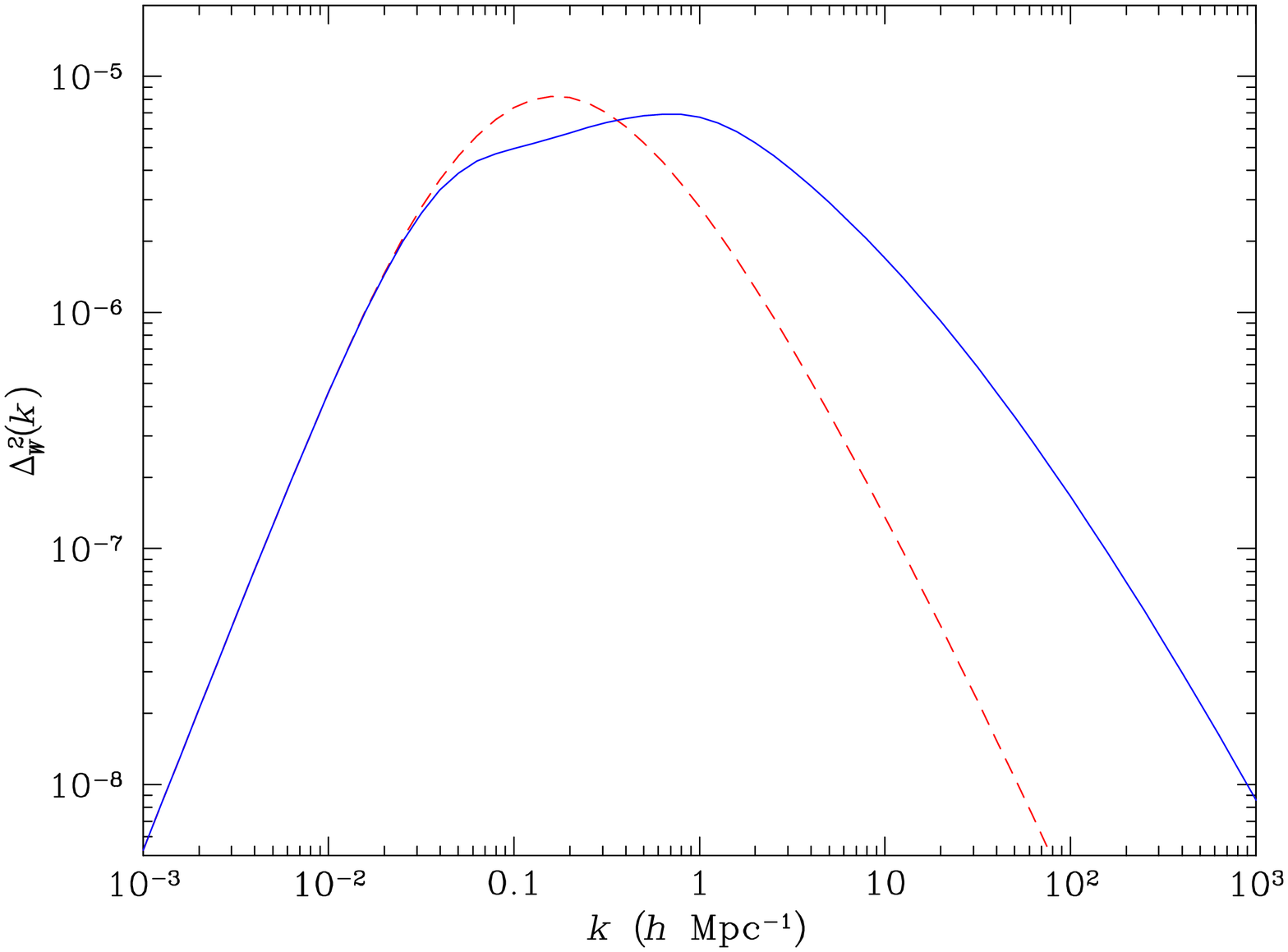}
%\null \vskip -0.4in
\figcaption{Spectral density of gravitational potential energy 
$ \Delta^2_W(k) $ [the integrand of eq.~(\ref{W})], 
evaluated at the present, plotted as a function of wavenumber $k$.
The dashed line shows $\Delta_W^2 $ in linear perturbation theory; 
the solid line shows the fully nonlinear form.
\label{P_W}}

Figure~\ref{P_W} shows the dimensionless spectral density of 
gravitational potential energy $ \Delta^2_W(k) $ defined in eq.~(\ref{W}), 
evaluated at the present, plotted as a function of wavenumber $k$.
The dashed curve shows the density in linear perturbation theory, 
and the solid curve shows its fully nonlinear form.

\includegraphics[width=\columnwidth]{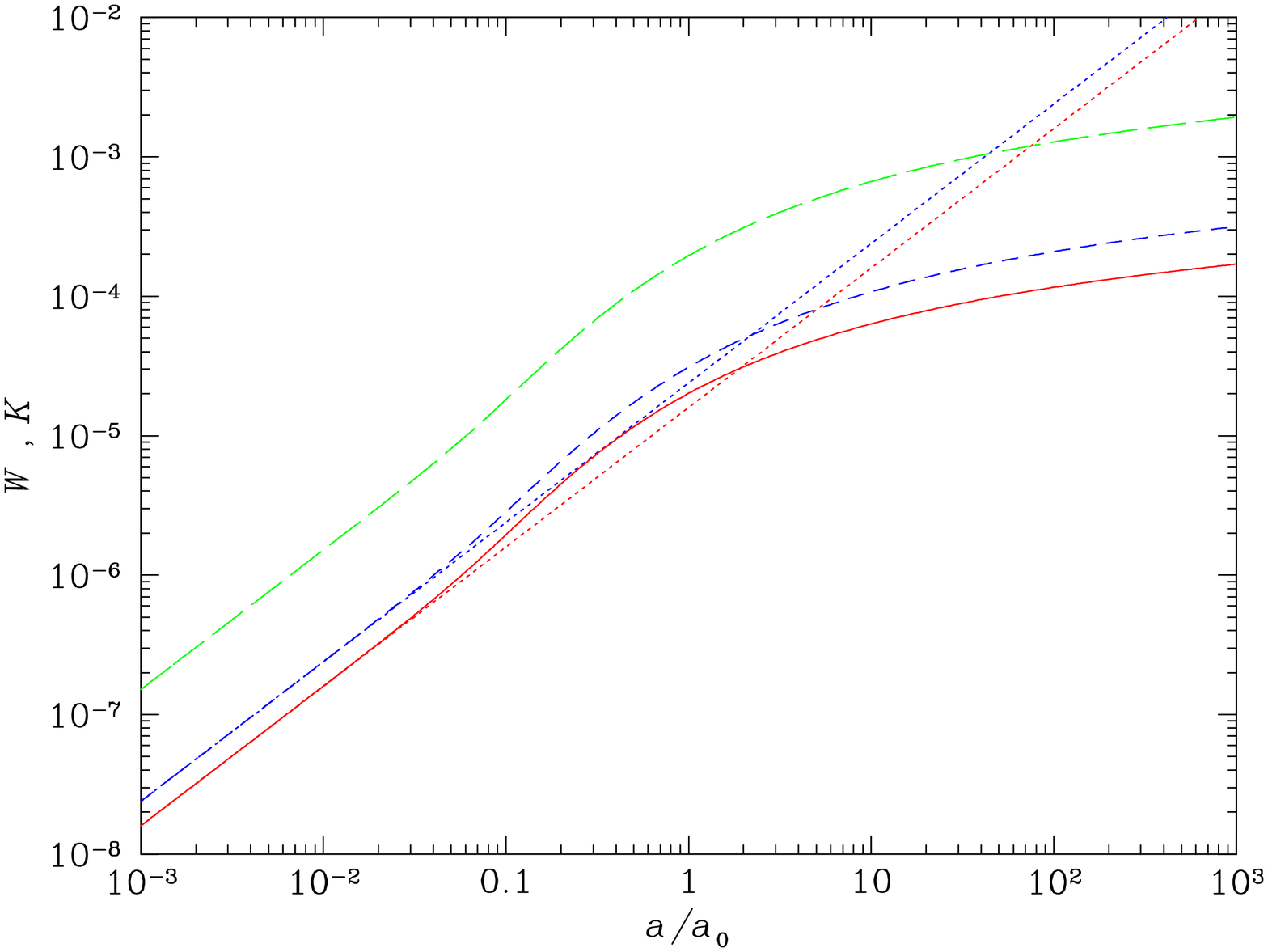}
%\null \vskip -0.4in
\figcaption{ Fractional contributions of gravitational potential energy 
$ W $ (long-dashed line) and kinetic energy $ K $ (solid line) 
to the total energy density of the universe, plotted as a function 
of past and future expansion factor for an $ \Omega_m = 1 $ universe.
The short-dashed line is the sum of contributions from inhomogeneities.
The dotted lines show results from linear perturbation theory.
\label{WK}}

Figure~\ref{WK} shows the contributions of potential energy 
and kinetic energy to the energy density of the universe, 
for past and future expansion factors in an $ \Omega_m = 1 $ universe.
At early times, perturbation theory gives an accurate result, 
but at $ a/a_0 \approx 0.05 $ (redshift $ z \approx 20 $) the behavior 
starts to change, for an interval growing faster than $a^1 $ 
with the fastest growth as $ a^{1.2} $, 
and then saturating and growing significantly more slowly, 
eventually as $\ln a$.

\section{Discussion}

In this {\it Letter} we have evaluated the size and the time evolution 
of the contribution of inhomogeneities to the expansion dynamics of 
a matter-dominated universe, 
including the effects of fully nonlinear density inhomogeneities.
When density fluctuations are in the linear regime, 
the ratio of the inhomogeneity contribution to the matter density grows 
linearly with expansion factor, as does curvature in an open universe, 
making only a very small contribution to the expansion rate.
As density fluctuations begin to go nonlinear, 
the inhomogeneity energy grows at a slightly faster rate, 
at most as $ a^{1.2} \propto a^{-3w} $, or $ w = -0.4 $.
This by itself, even if the dominant energy component, 
would be only temporarily and only very slightly accelerating, 
with deceleration parameter $ q_0 = {1 \over 2} (1 + 3w) = -0.1 $.
Since, at this time, the total fraction of inhomogeneity energy is 
$ \Omega_U \approx 10^{-5} \ll 1 $, this has a negligible effect on  
cosmological expansion dynamics.

As the universe further evolves, so that the main contributions 
to $W$ and $K$ come from deeply nonlinear scales, we compute the 
potential energy from integration of the nonlinear power spectrum, 
and obtain kinetic energy from the cosmic energy equation (eq.~[\ref{CEE}]).
In a scale-invariant model with power spectrum $ P \sim k^n $ 
as $ k \to 0 $, the kinetic and potential energies 
$K$ and $W$ scale with the expansion factor as $ a^{(1-n)/(3+n)} $ 
\citep{DP77} (logarithmically in $a$ as $ n \to 1 $), 
with ratio  
\begin{equation}
\label{ratio}
\frac{K}{|W|} = \frac{4}{7+n} \mathrm{.}
\end{equation}
Numerical simulations show that this continues to hold for the CDM spectrum 
with effective index $ n = d \log P / d \log k $ at an appropriate scale, 
the basis of the linear-nonlinear mapping \citep{PD94,PD96}.
For the CDM spectrum, with $ n \to 1 $ on large scales, 
this means that growth stops, and the ratio tends to the virial value 
$ K/|W| \to \frac12 $ at late times.
We note that aside from the integration of the Layzer-Irvine equation, 
many of these results were obtained by \citet{SH96}.

Our results show that the contributions of the potential and 
kinetic energies of inhomogeneities has never been strong enough 
to dominate the expansion dynamics of the universe.
For a universe with $ \Omega_m = 1 $ today, normalized 
to the large scale fluctuations in the microwave background, 
the net effect of inhomogeneities today is that of a slightly open  
universe, with $ \Omega_k \approx 10^{-4} $ in curvature.
The maximum contribution comes from scales of order 1 Mpc, 
falling off rapidly for smaller and larger $k$, as illustrated in 
Figure~\ref{P_W}.  
The behavior on asymptotically small scales ($k \gg 10^6 \, h \, 
{\rm Mpc}^{-1}$) depends on an extrapolation that 
ignores such details as star formation, but \citet{FP04} 
estimate that the net contribution of dissipative 
gravitational settling from baryon-dominated parts of galaxies, 
including main sequence stars and substellar objects, white dwarfs, 
neutron stars, stellar mass black holes, and galactic nuclei, 
is in total $ 10^{-4.9} $ of the critical energy density.

The suggestion that nonlinear effects for large inhomogeneities may mimic 
the effect of dark energy is not the case for the fully nonlinear theory.
It is true that higher order terms in perturbation theory grow faster; 
the general $n$-th order term grows as $ D^n(t) $.
There indeed comes a scale in space or an evolution in time 
where the behavior of higher order terms appears to diverge.
Nevertheless, the fully nonlinear result is well behaved.
It is only the perturbation expansion that breaks down, and 
the actual energy saturates and grows more and more slowly at late times.
As illustrated in Figure~\ref{WK}, the nonlinear potential and kinetic 
energies remain small compared to the total matter density at all times, 
even an expansion factor of $ 10^3 $ into the future.
Inhomogeneity effects do not substantially affect the expansion 
rate at any epoch.

\includegraphics[width=\columnwidth]{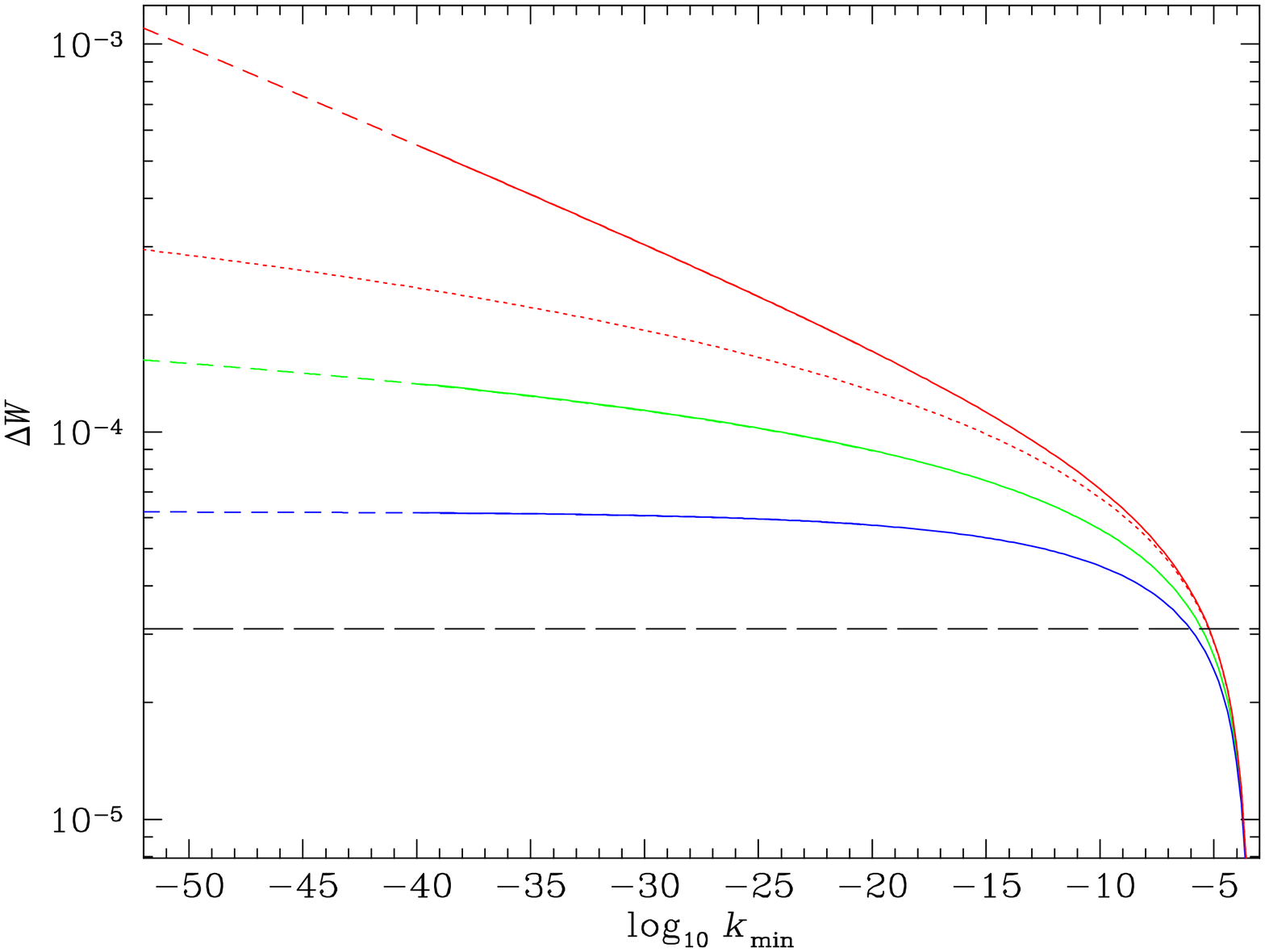}
%\null \vskip -0.4in
\figcaption{The expected fluctuation in the potential energy 
per unit mass $ \lexp (\Delta W)^2 \rexp^{1/2} $ evaluated at the present 
as a function of infrared cutoff $ k_\min $ for 
$ n = 0.95 $, $ n = 1 $, and $ n = 1.05 $ (solid lines, top to bottom).
Dashed lines are analytic approximations that asymptotically 
become $ k^{-0.025} $, $ (\log k)^{1/2} $, or constant, respectively.
The dotted line shows the result for a rolling spectral index 
that has $ n = 0.95 $ on the horizon today but approaches 
$ n = 1 $ as $ k \to 0 $, as predicted by most models of slow-roll inflation.
The mean value $ \lexp W \rexp = 3.1 \times 10^{-5} $ is shown 
as the horizontal dashed line.
\label{DW}}

It has been pointed out that although the average inhomogeneity 
energy is small, its variance has a logarithmically divergent 
contribution from the variance of the potential on 
super-horizon scales \citep{Kolb05a}, 
\begin{eqnarray} \label{var}
\lexp (\Delta W)^2 \rexp &=& {1 \over V^2} \int d^3x \, d^3x' \, 
{1 \over 4} \, \rhobar \lexp \phi(\x) \, \phi(\x') \rexp  \nonumber \\
&=& (2 \pi G \rhobar a^2)^2 \int {d^3k \over (2\pi)^3} \, 
{P(k) \over k^4} \, W^2(kR) , 
\end{eqnarray}
windowed over the horizon volume (for calculational convenience 
we use a Gaussian rolloff rather than a sharp radial edge).
For $ n \to 1 $ as $ k \to 0 $, this is indeed logarithmically 
dependent on the low-$k$ cutoff (and if $ n < 1 $ the divergence is worse), 
but the rest of the integral is finite for the CDM spectrum.
The fluctuation in potential energy, $ \lexp (\Delta W)^2 \rexp^{1/2} $, 
is shown in Figure~\ref{DW} as a function of the infrared cutoff $ k_\min $.
The integral is dominated by the smallest values of $k$, where 
perturbations are deep in the linear regime.
For $ n = 1 $ the result is very accurately 
$ \Delta W = 1.45 \times 10^{-5} \, | \ln k_\min R_H |^{1/2} $.
(We note that for $n \to 1$ the units of $ k_\min $ are unimportant.)
The fluctuation is comparable to the mean 
$ \lexp W \rexp = 3.1 \times 10^{-5} $ when the cutoff 
is near the scale of the horizon $ k = H_0/c $,  
and does not become of order 1 until 
$ k_\min \sim 10^{-170} $ (for $ n = 0.95 $), 
or $ k_\min \sim 10^{-10^9} $ (for $ n \to 1 $), or ever (for $ n > 1 $).
While such an exponentially vast range of scales may not be beyond the 
range of possibility in an inflationary universe, it requires a fearless 
extrapolation well beyond what is known directly from observation.
The fluctuation $ \Delta W $ is dominated by contributions from 
modes that are deep in the linear perturbation regime, and scales with 
expansion factor as $ \Delta W \propto \rhobar a^2 D $, constant in time.
This contribution to the energy will appear dynamically in the 
Friedmann equation as another matter component.  Furthermore, in the 
presence of a true dark energy component, any effects on cosmological 
expansion arising from inhomogeneities quickly becomes unimportant 
once dark energy becomes dominant \citep{SH96}.

%You may decide whether to include the following paragraph or not.

The fact that fluctuations in the potential diverge remains troublesome.
It has been recognized for some time that potential fluctuations 
in the standard model with $ n \to 1 $ are logarithmically divergent, 
but since for most purposes the value of the potential is unimportant, 
this has not been perceived as a significant problem.
The effect of potential on the expansion dynamics is real, 
but the weak logarithmic divergence and the fact that it is a 
feedback of a gravitational energy on gravitational dynamics may 
lead one to hope that this divergence is alleviated in a renormalized 
quantum theory of gravity. 

We have found that, to leading order in $ \phi $ and $ v^2 $ 
but with fully nonlinear density fluctuations, %on average 
inhomogeneities on sub-horizon scales have only a minimal effect 
on the cosmological expansion dynamics, even far into the future, 
and in particular never result in an accelerated expansion.
Other authors have also shown that recent attempts to explain an accelerated 
expansion through super-horizon perturbations face significant difficulties 
\citep{Flan05,Gesh05,Hira05}.
The possibility that a known component of the universe may be 
responsible for the accelerated expansion remains intriguing.
However, we conclude that sub-horizon perturbations are not a viable 
candidate for explaining the accelerated expansion of the universe.

\acknowledgements

We thank Dan Chung for helpful discussions on the general topic of
cosmological inhomogeneities and Uro\v{s} Seljak for pointing out 
the Seljak \& Hui reference and for helpful comments.
E.R.S. acknowledges support from the University of 
Florida's Alumni Fellowship program.
This research has made use of NASA's Astrophysics Data System.

%\clearpage


\begin{thebibliography}{}

\bibitem[Barausse et al.(2005)]{Bara05} 
Barausse, E., Matarrese, S., \& Riotto, A.\ 2005, 
%ArXiv Astrophysics e-prints, 
astro-ph/0501152 
%Inhomogeneities change the luminosity distance-redshift relation

\bibitem[Bardeen et al.(1986)]{BBKS86}
Bardeen, J.~M., Bond, J.~R., Kaiser, N., \& Szalay, A.~S.\ 1986, \apj, 304, 15

\bibitem[Bennett et al.(2003)]{Benn03}
Bennett, C.~L., et al.\ 2003, \apjs, 148, 1
%WMAP year 1 results

\bibitem[Bunn \& White(1997)]{BW97} Bunn, E.~F., \& White, 
M.\ 1997, \apj, 480, 6 
%COBE normalized \delta_H = 1.9 * 10^{-5}

\bibitem[Davis \& Peebles(1977)]{DP77}
Davis, M., \& Peebles, P.~J.~E.\ 1977, \apjs, 34, 425 
%On the integration of the BBGKY equations for the development 
% of strongly nonlinear clustering in an expanding universe

\bibitem[Flanagan(2005)]{Flan05}
Flanagan, \`E.~\`E.\ 2005, 
%ArXiv High Energy Physics - Theory e-prints, 
hep-th/0503202 
%Eanna Flanagan's retort to Kolb's claim

%\bibitem[Freedman et al.(2001)]{Free01}
%Freedman, W.~L., et al.\ 2001, \apj, 553, 47 
%%Confirm Hubble Expansion

\bibitem[Fukugita \& Peebles(2004)]{FP04}
Fukugita, M., \&  Peebles, P.~J.~E.\ 2004, \apj, 616, 643
% The Cosmic Energy Inventory   K+W = U = -7.7e-7 

\bibitem[Geshnizjani et al.(2005)]{Gesh05}
Geshnizjani, G., Chung, D.~J.~H., \& Afshordi, N.\ 2005, 
%ArXiv Astrophysics e-prints, 
astro-ph/0503553 
%Dan Chung's excellent response to Kolb's claim

%\bibitem[Hamilton et al.(1991)]{HKLM91}
%Hamilton, A.~J.~S., Kumar, P., Lu, E., \& Matthews, A.\ 1991, \apjl, 374, L1 
%% Reconstructing the primordial spectrum of fluctuations of the universe 
%%  from the observed nonlinear clustering of galaxies

\bibitem[Hirata \& Seljak(2005)]{Hira05}
Hirata, C.~M., \& Seljak, U.\ 2005, 
%ArXiv Astrophysics e-prints, 
astro-ph/0503582 
%Seljak's retort to Kolb's claim

\bibitem[Irvine(1961)]{Irv61} 
Irvine, W. M. 1961, Ph.D. thesis, Harvard University
%Local Irregularities in a Universe Satisfying the Cosmological Principle

\bibitem[Kolb et al.(2004)]{Kolb04} 
Kolb, E.~W., Matarrese, S., Notari, A., \& Riotto, A.\ 2004, 
%ArXiv Astrophysics e-prints, 
astro-ph/0410541
% Cosmological influence of super-Hubble perturbations

\bibitem[Kolb et al.(2005a)]{Kolb05a}
Kolb, E.~W., Matarrese, S., Notari, A., \& Riotto, A.\ 2005, \prd, 71, 023524 
% hep-ph/0409038
% Effect of inhomogeneities on the expansion rate of the Universe

\bibitem[Kolb et al.(2005b)]{Kolb05b}
Kolb, E.~W., Matarrese, S., Notari, A., \& Riotto, A.\ 2005, 
%ArXiv High Energy Physics - Theory e-prints, 
hep-th/0503117 
% Primordial inflation explains why the universe is accelerating today
% PRL version -- inflation explains why universe is accelerating

\bibitem[Layzer(1963)]{Layz63}
Layzer, D.\ 1963, \apj, 138, 174 
%Irving-Layzer (Cosmic Energy) Equation

\bibitem[Notari(2005)]{Nota05} 
Notari, A.\ 2005, 
%ArXiv Astrophysics e-prints, 
astro-ph/0503715 
%Subhorizon perturbations (Failure of Friedmann Equation [crazy])

\bibitem[Peacock \& Dodds(1994)]{PD94}
Peacock, J.~A., \& Dodds, S.~J.\ 1994, \mnras, 267, 1020 
%Reconstructing the complete Linear Power Spectrum

\bibitem[Peacock \& Dodds(1996)]{PD96}
Peacock, J.~A., \& Dodds, S.~J.\ 1996, \mnras, 280, L19 
%Reconstructing the complete non-linear Power Spectrum

\bibitem[Peebles(1980)]{Peeb80}
Peebles, P.~J.~E. 1980, The large-scale structure of the universe
(Princeton: Princeton University Press)
%The Book

\bibitem[Peebles \& Ratra(2003)]{Peeb03}
Peebles, P.~J.~E., \&  Ratra, B.\ 2003, Reviews of Modern Physics, 75, 559 
%Review of Dark Energy/Quintessence

\bibitem[Rasanen(2005)]{Ras05} Rasanen, S.\ 2005, ArXiv 
Astrophysics e-prints, arXiv:astro-ph/0504005 

\bibitem[Riess(2000)]{Ries00} 
Riess, A.~G.\ 2000, \pasp, 112, 1284 
%Riess's paper on SNIa, synthesizes his results and Perlmutter's results

\bibitem[Seljak \& Hui(1996)]{SH96}
Seljak, U., \& Hui, L.\ 1996, ASP Conf.~Ser.~88: 
Clusters, Lensing, and the Future of the Universe, 88, 267 
 

\end{thebibliography}
\end{document}